# MRXCAT-CDTI – A Numerical Cardiac Diffusion Tensor Imaging Phantom

Robbert J. H. van Gorkum[1], Jonathan L. Weine[1], William P. Segars[2], Christian T. Stoeck[1], Sebastian Kozerke[1].

[1]Institute for Biomedical Engineering, University and ETH Zurich, Zurich, Switzerland.

[2]Department of Radiology, Carl E Ravin Advanced Imaging Laboratories, The Duke University Medical Center, Durham, USA.

**Address for Correspondence:**

Sebastian Kozerke, PhD

Institute for Biomedical Engineering

University and ETH Zurich

Gloriastrasse 35

8092 Zurich

Tel.:   + 41 44 632 3549

Email:  kozerke@biomed.ee.ethz.ch

**Keywords:** Simulation, MRXCAT, cardiac diffusion tensor imaging, cardiac intravoxel incoherent motion

# Abstract

Magnetic Resonance cardiac diffusion tensor imaging (cDTI) and cardiac intravoxel incoherent motion imaging enables probing of in vivo myofiber architecture and myocardial perfusion surrogates. To study the impact of experimental parameters such as resolution, off-resonances and heart-rate variations, we propose a numerical open-source framework called MRXCAT-CDTI. It allows simulating diffusion and perfusion contrast for spin-echo (SE) and stimulated echo acquisition mode (STEAM) cDTI sequences. The Fourier encoder supports in-plane and/or through-slice off-resonance effects, as well as $T_2^*$ effects during single-shot image encoding. Optional lesions are included to mimic ischemic and infarcted myocardial regions. MRXCAT-CDTI allows assessing realistic influences on data acquisition, and how these affect the data encoding process and subsequent data processing. As an example, heart-rate variations lead to differences in partial saturation and relaxation of magnetization that end up in errors of 9 to 30% for cDTI angle metrics if not accounted for. For SE echo-planar cDTI, in-plane off-resonance effects more adversely affect cDTI metrics compared to through-slice off-resonances. With this work we propose an open-source MRXCAT-CDTI numerical simulation framework that offers realistic image encoding effects found in cardiac diffusion and perfusion data to systematically study influences of data encoding, reconstruction, and post-processing to promote reproducible research.

**Keywords (10):** Magnetic Resonance, Cardiac diffusion tensor imaging, Numerical simulation, MRXCAT.

# Introduction

Over the years, the interest in Magnetic Resonance (MR) cardiac diffusion imaging methods such as cardiac Diffusion Tensor Imaging (cDTI) [1–8] and cardiac intravoxel incoherent motion (cIVIM) imaging [9–12] has increased significantly. These techniques have helped to understand the consequence of diseases on the myocardial microstructure and consequently on the heart's function [2,3,5,8,13,14].

An important aspect of basic imaging research is to study the processes that underpin diffusion-related signal generation, such as the interaction of diffusing molecules carrying transverse magnetization with encoding gradients as well as how signal acquisition in $k$-space is affected by readout trajectories, $T_2^*$ decay, magnetic susceptibility-related effects and eddy currents.

Current cardiac diffusion simulation tools are mainly focused on the correspondence of diffusion contrast with the underlying microstructure. To this end, Monte Carlo (MC) simulations have been used to approximate the probability density distributions of material point displacements to study the effects of tissue structure on signal modulation, or to assess the sensitivity of cardiac diffusion sequences to myocardial perfusion [9,15–19]. Bulk motion suppression and intravoxel phase dispersion have been evaluated mainly for motion-compensated spin-echo (SE) sequences using cardiac tissue deformation information [20–26]. Using such methods, it has been shown that second-order motion compensation is essential for reproducible in vivo SE cDTI.

Besides material-point based MC simulations, explicit parametrized models have been used to generate diffusion signals for simplified geometric models [8,27–29]. Since anatomical cDTI models at arbitrarily high spatial resolution can be generated this way, they can serve as input for simulations of image acquisition and reconstruction algorithms. Thereby, the accuracy and precision of the imaging process itself can be studied and compared to the high-resolution ground truth data e.g. biases in diffusion metrics due to partial volume effects in the in-plane and through-slice directions with anisotropic acquisition resolutions. Moreover, accelerated acquisition strategies such as multi-band imaging [30,31] or compressed sensing [32–35] can be assessed. To date, the performance of acceleration techniques has been studied on prospectively or retrospectively accelerated in vivo acquisitions. A key limitation with using in vivo data is the lack of a ground truth.

Data encoding of cDTI sequences is known to be affected by, but not limited to, $T_2^*$ decay, field inhomogeneities and eddy current effects [28,36–39]. Understanding such influences on data encoding may be challenging in the in vivo cDTI [27,40] setting given the inherently low signal-to-noise (SNR). Moreover, any data acquisition process is subject to systematic (e.g. field inhomogeneities, $T_2^*$ decay), and random errors (e.g. noise, signal dropouts due to bulk motion, phase inconsistencies). To this end, a synthetic simulation framework can provide ground truth data to be able to determine how a process during data acquisition or reconstruction is affected.

MRXCAT [41] has been proposed as a hybrid numerical cardiovascular MR phantom designed based on the extended Cardiac-Torso (XCAT) phantom [42]. It provides synthetic ground truth data, which can be manipulated accordingly to optimize acquisition and reconstruction approaches. The XCAT phantom offers non-rigid respiratory and cardiac motion, and the use of non-uniform rational B-splines allows the object to be rasterized into arbitrarily fine spatial resolutions.

The objective of the present work is to provide cDTI and cIVIM extensions to the MRXCAT numerical phantom, referred to as MRXCAT-CDTI, to enable systematic optimization of cDTI and cIVIM image acquisition, reconstruction and inference. The open-source extension generates free-breathing, breath-held or static diffusion and perfusion tensor data, inclusion of off-resonance effects (in-plane and/or through-slice) and $T_2^*$ decay during echo-planar or non-Cartesian readouts, longitudinal magnetization ($M_z$) variations (i.e. partial saturation and relaxation) due to R-R variations, and a lesion model. Several MRXCAT-CDTI applications are showcased, providing image reconstruction and post-processing examples.

# Materials and Methods

The MRXCAT-CDTI phantom generates single-shot multi-coil $k$-space data $D(\vec{k}(t_s), n)$ for heartbeat $n$ according to following operator formalism

$$D(\vec{k}(t_s), n) = N \circ F(T_2^*) \circ B(\vec{x}) \circ S(N_c, r_c, \vec{x}_c) \circ C_d(b, \vec{g}) \circ C_m(T_E, T_{R,n}, \alpha) \\ \circ T(f, \bm{D}, \bm{D}^*, T_1, T_2, \rho) \circ O(\vec{x}, n), \quad (1)$$

where $O(\vec{x}, n)$ is the high-resolution spatial XCAT object for spatial positions $\vec{x} = [x, y, z]^T$ for a given heartbeat $n$ (including the option for breathing motion states). Function composition operator $\circ$ denotes the chaining of operators which result in $k$-space data $D(\vec{k}(t_s), n)$ for all $k$-space sampling time points $t_s$. The order of application of operators is from right to left. The tissue operator $T$ maps relaxation times $T_1(\vec{x})$, $T_2(\vec{x})$, as well as proton density $\rho(\vec{x})$, cIVIM perfusion fraction $f(\vec{x})$, diffusion tensor $\bm{D}(\vec{x})$ and perfusion (pseudo-diffusion) tensor $\bm{D}^*(\vec{x})$ to the object $O(\vec{x}, n)$. The sequence operator $C_m$ defines signal model $m$, e.g. SE or stimulated echo acquisition mode (STEAM), parametrized by $T_E, T_{R,n}$, and excitation angle $\alpha$ to consider $B_1^+$ inhomogeneity and slice profile effects. The diffusion contrast operator $C_d$ simulates the application of diffusion encoding gradients by normalized diffusion encoding directions $\vec{g}$ and the $b$-value. Coil sensitivities are defined by coil operator $S$, for $N_c$ receive coils with radius $r_c$ at spatial locations $\vec{x}_c$. Off-resonance operator $B$ adds spatially dependent off-resonance effects in both in-plane and through-slice directions. Operator $F$ defines general Fourier encoding including non-uniform sampling with $T_2^*(\vec{x})$ weighting and off-resonance $B(\vec{x})$ for the multi-coil $k$-space data. Complex-valued zero-mean normal noise with standard deviation $\sigma$ is added to the $k$-space data by operator $N$. A flow diagram of the MRXCAT-CDTI software structure is presented in Figure 1.

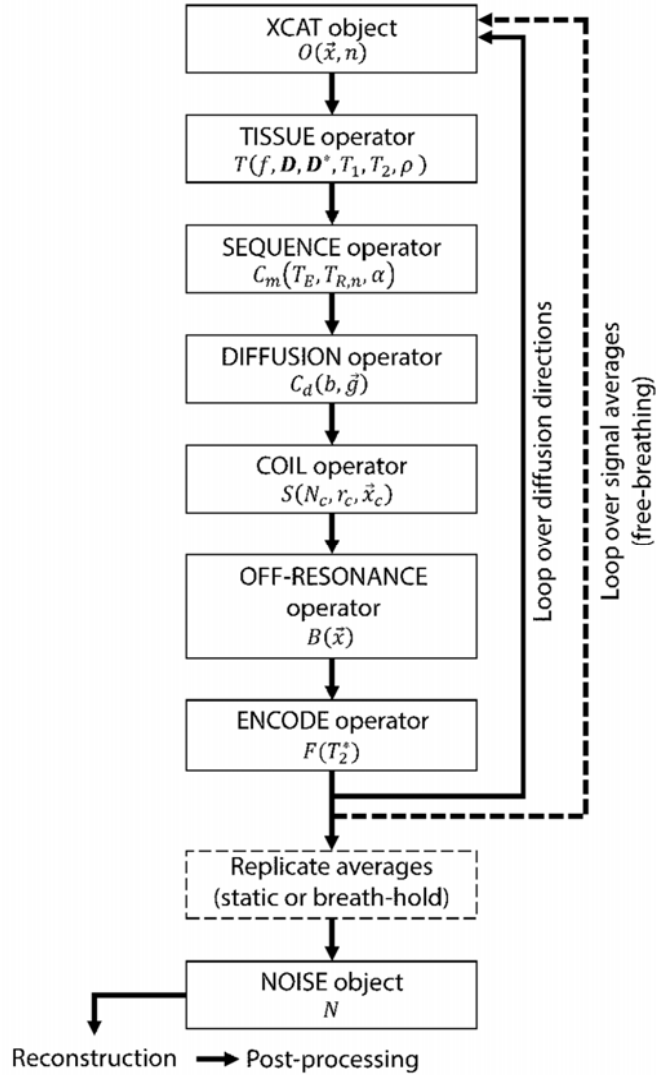

**Figure 1.** Flow diagram describing MRXCAT software and workflow structure. The tissue masks of the XCAT object at heartbeat $n$ are mapped with tissue, diffusion, perfusion, and lesion properties. Sequence contrast is then added with the sequence operator, and the diffusion operator subsequently adds the corresponding MR diffusion and perfusion contrast. After multiplication with coil sensitivities the data can then be Fourier encoded in the presence of $T_2^*$ maps and field maps on a coil-per-coil basis. The process is repeated for each diffusion direction and, alternatively, signal averages, before noise is added to obtain the user-defined SNR. The data is then ready for further reconstruction and post-processing tasks.

## Tissue operator

Tissue properties $T_1(\vec{x})$, $T_2(\vec{x})$, and proton density $\rho(\vec{x})$ are mapped using the methodology described in [41], with the addition of a normal distributed tissue parameter $T_2^*(\vec{x})$ having a user-defined mean and standard deviation. The extension presented in this work includes the generation of diffusion and perfusion tensors for the left ventricle, modeling the structure of the myocardium as Gaussian compartments. The tensors maps are constructed by log-Euclidean interpolation of randomly sampled tensors. The tensor distributions reflect the statistics of in vivo findings for the implemented sequences.

In MRXCAT-CDTI tensors can either be defined locally from a default eigen system without spatial variation of eigenvalues over the left ventricle, or by using random eigenvalue maps generated with a rejection sampling approach. Both methods utilize the high-resolution left-ventricular (LV) XCAT organ masks to generate the eigenvalue/tensor maps.

The rejection sampling was implemented in Python (version 3.6.9, Python Software Foundation, https://www.python.org/) and TensorFlow (version 2.1, https://www.tensorflow.org/). It yields a two-dimensional uniform distribution for fractional anisotropy (FA) and the tensor trace over defined intervals, while also constraining the eigenvalues to defined intervals. Parameter supports for diffusion and perfusion tensors for both healthy and lesioned regions can be found in Supplementary Materials Table S1.

In brief, the sampling strategy is as follows: First, a tensor-trace value is uniformly sampled from the given interval. Second, a triple of eigenvalues whose sum is equal to the previously obtained trace value is sampled from the given intervals. In the third and last step, the obtained eigenvalue triple is rejected based on a MC decision with respect to the value. The probabilities used in the MC step are calculated such that the resulting distribution over FA (after rejection) is approximately uniform. Non-rejected distributions are burned in using $3 \cdot 10^5$ samples.

Generating maps containing eigenvalues inside the XCAT LV masks from the previously generated eigenvalue samples is achieved by tensor interpolation in log-Euclidean space [43]. The interpolation uses a radial basis function kernel and geodesic distance in polar coordinates. The reference tensors for interpolation are constructed from a helix angle (HA) range (−60° to +60° linearly varying over the transmural depth from epi- to endocardium) and an absolute second eigenvector (absE2A) sheetlet angle [13] depending on the position of the reference point inside the LV mask. A random eigenvalue triple is used to scale the tensor accordingly. The average distance between reference tensors defines the scale on which the resulting map is smooth. The absE2A map is randomly generated prior to log-Euclidean interpolation also using interpolation of a scalar angle inside the same LV mask.

Sequence operator

Transverse magnetization for the SE sequence $C_{SE}$ is calculated recursively according to

$$C_{SE}(T_E, T_{R,n+1}, \alpha) = \left( M_0 \left( 1 - e^{-\frac{T_{R,n+1}}{T_1(\vec{x})}} \right) + M_z(T_{R,n}) \cos(\alpha) e^{-\frac{T_{R,n+1}}{T_1(\vec{x})}} \right) \sin(\alpha) e^{-\frac{T_E}{T_2(\vec{x})}}, \quad (2)$$

with $M_0$ the initial (full) magnetization, $T_{R,n+1}$ the R-R interval duration of heartbeat $n+1$, $T_1(\vec{x})$ the local $T_1$ value, $M_z$ the (partially) recovered longitudinal magnetization, $T_{R,n}$ the R-R interval duration of heartbeat $n$, $\alpha$ the flip angle, $T_E$ the echo time, and $T_2(\vec{x})$ the local $T_2$ value.

For the STEAM sequence, transverse magnetization $C_{STEAM}$ is generated recursively according to

$$C_{STEAM}(T_E, T_{R,n+2}, \alpha) = \frac{1}{2} \left( M_0 \left( 1 - e^{-\frac{T_{R,n+1}}{T_1(\vec{x})}} \right) + M_z(T_{R,n}) \cos(\alpha) e^{-\frac{T_{R,n+1}}{T_1(\vec{x})}} \right) e^{-\frac{T_{R,n+2}}{T_1(\vec{x})}} \sin(\alpha) e^{-\frac{T_E}{T_2(\vec{x})}}, \quad (3)$$

where $T_{R,n+2}$ is the R-R interval duration of heartbeat $n+2$, while $T_E \ll T_1(\vec{x})$ is assumed.

Diffusion and perfusion contrast operator

Diffusion contrast operator $C_d$ simulates LV myocardium signal contrast using the diffusion and perfusion tensors according to

$$C_d(b, \vec{g}) = f(\vec{x}) \cdot e^{-b\vec{g}^T \boldsymbol{D}^*(\vec{x})\vec{g}} + \left(1 - f(\vec{x})\right) \cdot e^{-b\vec{g}^T \boldsymbol{D}(\vec{x})\vec{g}}, \qquad (4)$$

with $f(\vec{x})$ being the perfusion fraction, $\vec{g}$ normalized diffusion encoding direction, $\boldsymbol{D}^*(\vec{x})$ the local perfusion tensor, $\boldsymbol{D}(\vec{x})$ the local diffusion tensor, and $b$ the sequence's $b$-value. Different diffusion and perfusion tensors can be assigned to SE and STEAM simulations.

Several diffusion encoding schemes are provided such as the scheme mentioned in [44] to reduce the blood-pool signal for in vivo acquisitions. Schemes with 6, 9, 12, 15, and 30 directions generated by the QMRTools [45] are additionally included. The latter schemes can be set up for both cDTI and cIVIM acquisitions simply by adjusting the $b$-value array.

In case of STEAM, the nominal $b$-values can be scaled with respect to the current R-R interval to mimic the influence of in vivo heart rates on the $b$-value. Assuming the diffusion gradients are applied instantaneously, the actual $b$-value is determined by

$$b_{act,n+2} = b_{nom} \cdot \frac{T_{R,n+2}}{T_{R,nom}}, \qquad (5)$$

with $b_{act,n+2}$ the actual $b$-value at index $n$, $b_{nom}$ the nominal $b$-value, $T_{R,n+2}$ the current R-R interval duration at index $n$, and $T_{R,nom}$ the nominal R-R interval duration.

Lesion operator

Lesions can be defined inside the LV myocardium as a function of sector width and transmural depth. This allows for flexibility when defining the circumferential and transmural extent of the lesions across the myocardium. An example lesion mask, perfusion fraction $f$ map and corresponding distributions can be seen in Supplementary Materials Figure S1. The user can assign a different perfusion fraction distribution in the lesion area. In addition, healthy and lesion tissue types can be assigned with different diffusion and perfusion tensors to be used in Equation 4.

## Encoding operator

Before inclusion of off-resonance effects and $T_2^*$ weighting during signal encoding, the transverse magnetization $\rho'$ can be described by the following relation

$$\rho'(\vec{x}, n) = C_d(b, \vec{g}) \circ C_m(T_E, T_{R,n}, \alpha) \circ T(f, \boldsymbol{D}, \boldsymbol{D}^*, T_1, T_2, \rho) \circ O(\vec{x}, n), \qquad (6)$$

with $C_d$ being the diffusion contrast operator, $b$ the $b$-value, $\vec{g}$ normalized diffusion encoding direction, $C_m$ the sequence operator, $T_E$ echo time, $T_{R,n}$ the R-R interval duration of heartbeat $n$, $\alpha$ the flip angle, $T$ the tissue operator, perfusion fraction $f$, diffusion tensor $\boldsymbol{D}$, perfusion tensor $\boldsymbol{D}^*$, spin-lattice relaxation time $T_1$, spin-spin relaxation time $T_2$, proton density $\rho$, $O$ the high-resolution XCAT object, $\vec{x} = [x, y, z]^T$, and $n$ the heartbeat index.

The simplified complex time-domain signal $D(\vec{k}(t_s), n)$ for $k$-space locations $\vec{k}$ at sampling times $t_s$ is then given by

$$D(\vec{k}(t_s), n) = N \circ F(T_2^*) \circ B(\vec{x}) \circ S(N_c, r_c, \vec{x}_c) \circ \rho'(\vec{x}, n), \qquad (7)$$

with $N$ being the noise operator, $F$ the Fourier encoder, $T_2^*$ the off-resonance affected $T_2$ relaxation time, $B$ the off-resonance operator, and $S$ the coil operator for $N_c$ the receive coils with radius $r_c$ at spatial locations $\vec{x}_c$.

A field map can be defined using the same XCAT tissue masks as for mapping the tissue properties. As a result of the presence of off-resonance field gradients, in e.g. EPI, the object can be either compressed or stretched depending on the EPI phase-encode blip direction [28,46,47]. The severity of image distortions using an EPI readout is dependent on the strength of the field gradient [Hz/pixel] with respect to the acquisition bandwidth (BW) [Hz/pixel] along the phase-encode ($k_y$) direction. In MRXCAT-CDTI, the field gradient in the LV myocardium can be adjusted to deliver the desired level of object distortion. Moreover, the user can define the target acquisition BW of the readout trajectory.

High resolution off-resonance maps $\Delta F_0(\vec{x})$ are generated from MRXCAT anatomical masks. Fatty liver signal can be set to -220 Hz or -440 Hz to mimic the chemical shift at 1.5T or 3T, respectively. A fat fraction parameter then adds a portion of the liver fat off-resonance signal as a frequency offset. The simulated field map allows for the inclusion of an analytically-defined off-resonance area in the vicinity of the posterior vein [48,49]. An initial field map is obtained by fitting a smoothing spline over the map components [50,51]. The field map can then be scaled such that the LV sector containing the off-resonance effect induced by the posterior vein matches the desired user-defined target gradient value. Alternatively, MRXCAT-CDTI provides an example of reformatting an in vivo field map to match the XCAT mask orientation. The reformatted in vivo field map can then be incorporated into the forward model simulation.

To mimic a through-slice off-resonance gradient, the field maps are replicated for the user-defined through-slice positions $dz$ adding a frequency offset according to $\Delta F_0(x, y, dz) = \Delta F_0(x, y) + \Delta F_{dz}(dz)$. The user can choose between linear or quadratic through-slice frequency distributions. An example field map and through-slice frequency profile is illustrated in Supplementary Materials Figure S2.

The encoding process projects the object onto the image grid with voxel volume $\Delta V$. In order for the discrete-to-discrete model to provide realistic encoding effects, the encoded object and other spatial maps have higher BW per spatial axis than the target resolution [52]. To balance simulation times and partial volume effects, the presented work uses a five-fold BW increase per spatial axis. This yields an acceptable rasterization error in the range of $10^{-2}$ to $3 \cdot 10^{-2}$ [52]. The encoder can include temporal $k$-space filtering caused by $T_2^*$ decay and the temporal phase evolution due to off-resonances. Additionally, the encoder supports non-uniform $k$-space sampling. For image-reconstruction, the non-uniform Fourier transform (NUFFT) is used [53,54], including a NUFFT wrapper [55,56].

Noise addition

The noise standard deviation is iteratively refined using the encoded images to reach a desired SNR ($SNR_{target}$) within 1% accuracy. The SNR is calculated after Roemer coil combination [57] by determining the mean LV signal and dividing it by the standard deviation of the real part of the complex noise vector. MRXCAT-CDTI additionally allows the user to generate noise based on the same random number generator seed for each dataset, which can assist in examining systematic and random effects in a controlled manner.

## In vivo and MRXCAT-CDTI data acquisition

*Multi-slice in vivo vs. MRXCAT-CDTI comparison*

In vivo second-order motion-compensated SE cDTI [21,24] data was acquired from one healthy volunteer (male, age 26, heart rate 58±5 bpm). Imaging was performed on a clinical 1.5T Philips Achieva System (Philips Healthcare, Best, The Netherlands) using a 32-channel cardiac array coil and a gradient system delivering 80 mT/m gradient amplitude at 100 T/m/s slew rate. The volunteer provided written informed consent according to institutional policy.

Imaging was performed in short-axis orientation at base, mid and apical levels, using a reduced field-of-view (FOV) technique [58,59] and spectral-spatial water excitation [60]. Imaging parameters were as follows: spatial resolution = 2.5x2.5x8 mm², FOV = 230x111 mm², signal averages = 10, $T_R$ = 3 R-R interval (R corresponds to the R-wave of the electrocardiogram (ECG)), $T_E$ = 88 ms, $BW_{epi}$/pixel = 23.4 Hz/pixel / echo-spacing = 1 ms, trigger delay (time between detection of R-wave and application of first RF pulse) = 65% of peak-systole, and a diffusion gradient scheme with 3 orthogonal directions at $b$ = 100 s/mm² and 9 directions at $b$ = 450 s/mm² [40]. Diffusion images were acquired during free-breathing, using respiratory navigation with a 7 mm gating window and slice tracking [10].

A corresponding short-axis image orientation MRXCAT-CDTI dataset at base, mid and apical levels was simulated with the following settings: motion states = 10, average SNR in myocardium for a single $b$ = 0 s/mm² image = 10, signal averages = 10, $T_R$/$T_E$ = 3000/88 ms, and $BW_{epi}$/pixel = 23.4 Hz/pixel / echo-spacing = 1 ms. A field map was simulated with gradient $\nabla F_y$ of 8 Hz/pixel at the inferolateral LV heart wall, fat frequency offset = -220 Hz, and fat fraction = 10%. An overview of the forward model field maps is provided in Supplementary Materials Figure S3. Each signal average was randomly assigned to one of 10 available motion state masks to mimic free-breathing respiratory-motion variations. Variations in $M_z$ due to excitation history were introduced by applying a standard deviation (SD) on the $T_R$ of 10%.

*MRXCAT-CDTI excitation history correction example*

A static single short-axis mid-ventricular slice was simulated with average SNR in myocardium for a single $b = 0$ s/mm² image = 20, signal averages = 10, and $T_R/T_E$ = 1000/88 ms. Variations in $M_z$ due to excitation history were introduced by applying a SD on the $T_R$ of 10%.

*MRXCAT-CDTI off-resonance example*

A static single short-axis mid-ventricular slice was simulated with average SNR in myocardium for a single $b = 0$ s/mm² image = 20, signal averages = 10, $T_R/T_E$ = 2000/83 ms, and $BW_{epi}$/pixel = 35 Hz/pixel / echo-spacing = 0.7 ms. A ground truth dataset was simulated with no off-resonance effects. A field map was simulated with a $\nabla F_y$ of 17.5 Hz/pixel at the posterior vein location, fat frequency offset = -440 Hz, and fat fraction = 10%. The field map was replicated 5-fold along the through-slice direction ($dz$), and a quadratic through-slice frequency profile with mean gradient strength of 7.5 Hz across the slice was added. The field map used in the forward model is shown in Supplementary Materials Figure S2.

*MRXCAT-CDTI cIVIM lesion example*

A single-slice short-axis mid-ventricular static image was simulated with average SNR in myocardium for a single $b = 0$ s/mm² image = 50, signal averages = 1, and a diffusion gradient scheme with 6 directions, with $b$-values 0, 25, 50, 75, 100, 150, 200, 250, 300, 400, and 500 s/mm². Perfusion fraction $f$ was set to 0.14±0.04 in the bulk of the myocardium, and to 0.01±0.01 in the lesion area.

A detailed overview of all MRXCAT-CDTI simulation parameters can be found in the Supplementary Materials.

*Computational performance*

MRXCAT-CDTI was written in MATLAB (9.7, MathWorks, Natick, MA, USA). All MRXCAT-CDTI data was simulated using 30 cores on a high-performance cluster. Computational performance was assessed for all MRXCAT-CDTI tasks by providing the simulation times in mean±SD.

## Data reconstruction

In vivo data was reconstructed with MRecon (GyroTools LLC, Winterthur, Switzerland) using 10 virtual coils [61]. A non-rigid registration was applied to multi-slice in vivo and MRXCAT-CDTI data before computing the diffusion tensors [62].

## Data analysis

*Multi-slice in vivo vs. MRXCAT-CDTI comparison*

In vivo R-R intervals were derived from the ECG recorded signal, and for MRXCAT-CDTI data the simulated R-R intervals were used. Excitation history correction was performed using the following correction factor $K$ at heartbeat index $n$. The correction scales the data to the heartbeat duration $T_{R,1}$ according to

$$K(n+1) = \frac{\left(M_0\left(1-e^{-\frac{T_{R,1}}{T_1(\vec{x})}}\right) + M_z(T_{R,0})\cos(\alpha)e^{-\frac{T_{R,1}}{T_1(\vec{x})}}\right)}{\left(M_0\left(1-e^{-\frac{T_{R,n+1}}{T_1(\vec{x})}}\right) + M_z(T_{R,n})\cos(\alpha)e^{-\frac{T_{R,n+1}}{T_1(\vec{x})}}\right)}, \quad (8)$$

with $M_0$ the initial (full) magnetization, $T_{R,n}$ the R-R interval duration of heartbeat $n$, $M_z$ the (partially) recovered longitudinal magnetization, $\alpha$ the flip angle, and $T_1(\vec{x})$ the spatially-dependent $T_1$ relaxation times.

When the flip angle is set to 90°, the equation is simplified to

$$K(n+1) = \frac{\left(1 - e^{-\frac{T_{R,1}}{T_1(\vec{x})}}\right)}{\left(1 - e^{-\frac{T_{R,n+1}}{T_1(\vec{x})}}\right)}. \tag{9}$$

A global myocardial $T_1$ value of 1000 ms was considered for in vivo and MRXCAT-CDTI datasets. Multiplication of data by the correction factor then yielded the $M_z$ corrected data. Equations for STEAM-based $M_z$ correction are given in the Supplementary Materials section.

Upon computation of the diffusion tensors [63], HA, transverse angle (TA), absE2A, mean diffusivity (MD) and FA were evaluated [13,63,64]. From LV ROI data histograms were computed to provide mean±standard deviation (SD) values for each cDTI metric and to compute normalized root-mean-square error (nRMSE) values relative to ground truth data for each cDTI metric according to $nRMSE(\varepsilon) = \sqrt{\|\varepsilon - \varepsilon_{ref}\|^2 / \|\varepsilon_{ref}\|^2}$. Only the inner 80% transmural depth data was considered for processing. For the helix angle, 10-20% was defined as the endocardial region, and 80-90% was defined as the epicardial region.

*MRXCAT-CDTI excitation history correction example*

$M_z$ correction was applied to SE cDTI data using Equation 9 prior to computing diffusion tensors, assuming a global myocardial $T_1$ value of 1000 ms. After determining the diffusion tensors, LV ROI data histograms were computed to provide mean±SD values for each cDTI metric and to compute nRMSE values relative to ground truth data for each cDTI metric. The transmural, endocardial and epicardial regions for processing were defined the same as for the *Multi-slice in vivo vs. MRXCAT-CDTI comparison*.

*MRXCAT-CDTI off-resonance example*

For each simulation case, a LV ROI was used to compute the structural similarity with respect to ground truth data and is reported using mean±SD [65]. After computing the diffusion tensors, the LV myocardium ROI was subdivided into six sectors for analysis. Data from the unaffected remote and off-resonance affected sectors were used to compute histograms for each cDTI metric, and to provide mean±SD values for the ROI. Histogram intersection values (*HI*) were computed from the normalized histograms of the LV

ROI data ($I$) for HA, TA, absE2A, MD, and FA with respect to the ground truth data ($M$), according to $HI(I,M) = \frac{\sum_{k=1}^{N_{bins}} \min(I_k, M_k)}{\sum_{k=1}^{N_{bins}} M_k}$ [66], with $N_{bins}$ = 35. Only the inner 80% transmural depth data was considered for processing. For off-resonance correction using MRXCAT-CDTI, the reader is referred to [28].

*MRXCAT-CDTI cIVIM lesion example*

Simulated cIVIM data was subjected to least-squares-fitting using a tensor model as described in [9] with a *b*-value split of 250 s/mm² and excluding the *b* = 0 s/mm² data, yielding spatial parameter maps related $\boldsymbol{D}(\vec{x})$ and $\boldsymbol{D}^*(\vec{x})$ (i.e. $MD(\vec{x})$, $FA(\vec{x})$, $MD^*(\vec{x})$, $FA^*(\vec{x})$ and $f(\vec{x})$). The myocardial LV ROI was divided into six sectors, and data from an unaffected remote zone was compared to the lesion area. Sector data was used to determine mean±SD values. Only the inner 80% transmural depth data was considered for processing.

# Results

## Multi-slice in vivo vs. MRXCAT-CDTI comparison

In Figure 2 a comparison between MRXCAT-CDTI simulation framework with in vivo cDTI data is shown including excitation history correction. Basal, mid-ventricular, and apical slices are displayed with mean diffusion weighted image (DWI) and cDTI metric maps. MRXCAT-CDTI mean DWI data exhibits similar global appearance, noise levels, and partial volume effects as the in vivo cDTI data, without the residual left and right ventricular lumen signal. In the presented example off-resonance gradients located in the inferolateral segment have been included in the simulation. Changes in TA are apparent for both in vivo and simulated data in the affected segments.

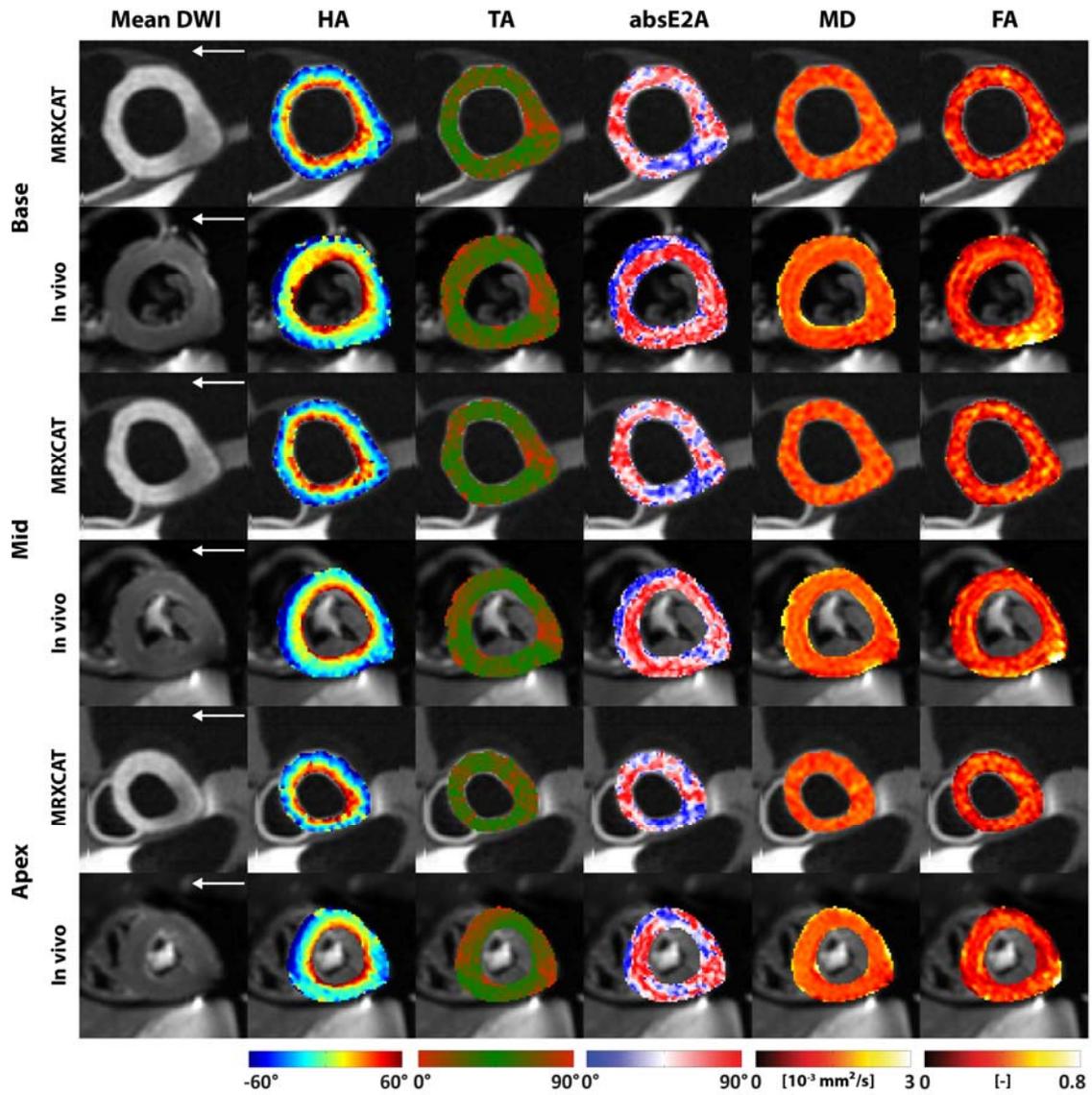

**Figure 2.** Example MRXCAT-CDTI and in vivo cDTI data corrected for the excitation history. Three base-to-apex short-axis slices are shown with corresponding cDTI metrics. White arrows indicate phase-encode direction, HA: helix angle, TA: transverse angle, absE2A: absolute sheetlet angle, MD: mean diffusivity, FA: fractional anisotropy.

A histogram overview of cDTI metrics for MRXCAT-CDTI and in vivo excitation history corrected data is shown in Figure 3 which displays comparable data distributions for TA, MD, and FA. Supplementary Materials Table S2 summarizes excitation history corrected data cDTI metrics, demonstrating the full potential of MRXCAT-CDTI to realistically mimic in vivo cDTI data distributions with comparable mean and SD values.

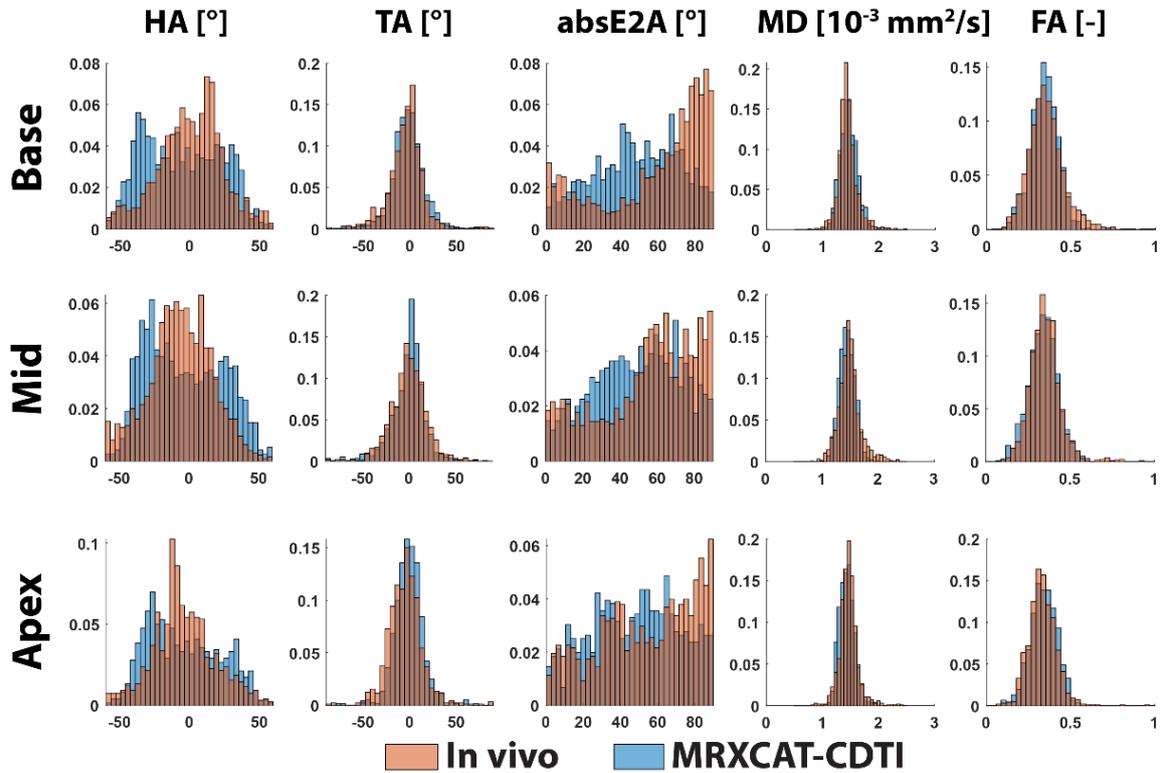

**Figure 3.** Example MRXCAT-CDTI and in vivo cDTI data corrected for the excitation history. Three base-to-apex short-axis slices are shown with corresponding cDTI metrics. White arrows indicate phase-encode direction, HA: helix angle, TA: transverse angle, absE2A: absolute sheetlet angle, MD: mean diffusivity, FA: fractional anisotropy.

The distribution of R-R intervals for the in vivo and MRXCAT-CDTI multi-slice datasets are shown in histogram form in Supplementary Materials Figure S4. The mean±SD values are 2.91±0.24 seconds for the in vivo dataset, and 2.99±0.31 seconds for the MRXCAT-CDTI dataset.

## MRXCAT-CDTI excitation history correction example

An overview of mean DWI and cDTI metric maps for ground truth, excitation history uncorrected and corrected data can be found in Figure 4. All data show comparable map quality, with no observable difference.

In Figure 5, the histograms of the ground truth, excitation history uncorrected and excitation history corrected cases are displayed. No noticeable differences between the cases and cDTI metrics can be observed. The corresponding mean±SD values of the distributions can be found in Supplementary Materials Table S3. All cases exhibit comparable mean and standard deviation for each cDTI metric.

Supplementary Materials Table S4 shows the nRMSE values of the single-slice MRXCAT-CDTI data for the excitation history corrected and excitation history uncorrected data. nRMSE values for excitation history uncorrected and excitation history corrected values are for the endocardial HA 3.6% and 0.8%, epicardial HA 11.4% and 5.1%, TA 30.9% and 6.7%, absE2A 9.5% and 1.9%, MD 1.3% and 0.2%, and FA 5.4% and 0.9%. The excitation history uncorrected data shows an average nRMSE of 10.3±10.7% across the cDTI metrics, whereas excitation history corrected data shows an nRMSE of 2.6±2.7% across all cDTI metrics, a 4.0-fold decrease.

The R-R interval distribution of the MRXCAT-CDTI single-slice dataset is shown in histogram form in Supplementary Materials Figure S4, with a mean±SD value of 1.01±0.12 seconds.

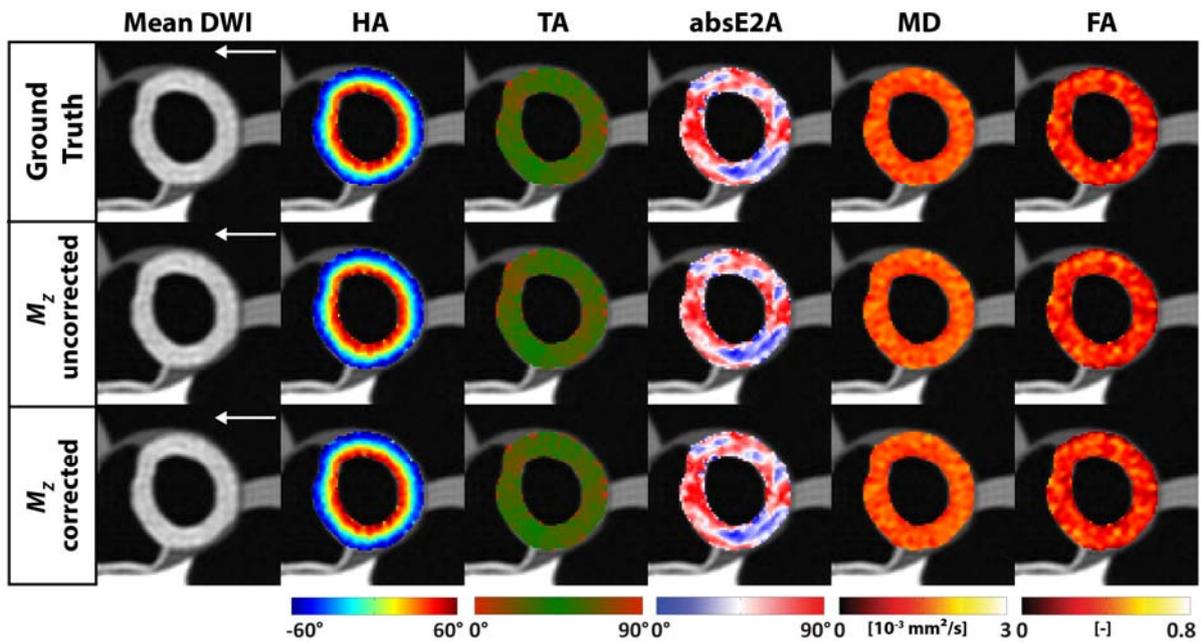

**Figure 4.** MRXCAT-CDTI single-slice example of partial saturation effects due to varying heart rate ($M_z$). Effects on cDTI metrics relative to ground truth (top row), data uncorrected for the excitation history (middle row), and data corrected for the excitation history (bottom row). White arrows indicate the phase-encode axis, HA: helix angle, TA: transverse angle, absE2A: absolute sheetlet angle, MD: mean diffusivity, FA: fractional anisotropy.

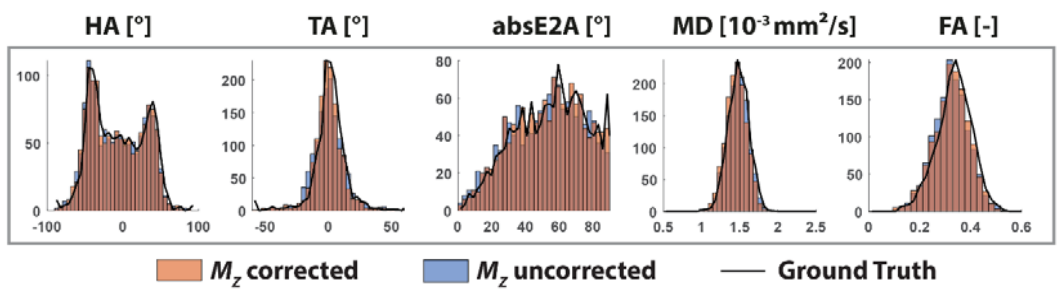

**Figure 5.** Histograms of excitation history corrected MRXCAT-CDTI single-slice data (from Figure 4) for excitation history uncorrected (blue), and corrected cases (red). Each histogram is plotted against ground truth data (solid black line).

## MRXCAT-CDTI off-resonance example

Off-resonance MRXCAT-CDTI examples are shown Figure 6. The introduction of through-slice gradients does not visually affect the appearances of the cDTI maps compared to ground truth data. Through-slice and in-plane gradients introduce a reduction of LV signal magnitude in the mean DWI in the off-resonance affected sector, as well as stretching of the posterior LV wall. When both in-plane and through-slice effects are present, differences in HA, TA, MD and FA can be observed. Structural similarity for the through-slice case was 99.9±0.1%, compared to 56.7±4.6% when both through-slice and in-plane off-resonance effects are present.

In Figure 7, histograms are shown for an off-resonance affected sector, and a remote sector. The through-slice case shows similar distributions compared to the ground truth distributions for both remote and affected zones. In the presence of both in-plane and through-slice effects, differences in HA, TA, absE2A, and MD are noticeable for the affected zone, whereas the remote zone stays unaffected. Supplementary Materials Table S5 displays an overview of each metrics per sector. Compared to ground truth and through-slice effects only, larger standard deviations are noticeable in the affected sector for the through-slice and in-plane off-resonance case for the endocardial HA, epicardial HA and TA. MD and FA values remain comparable across sectors and cases.

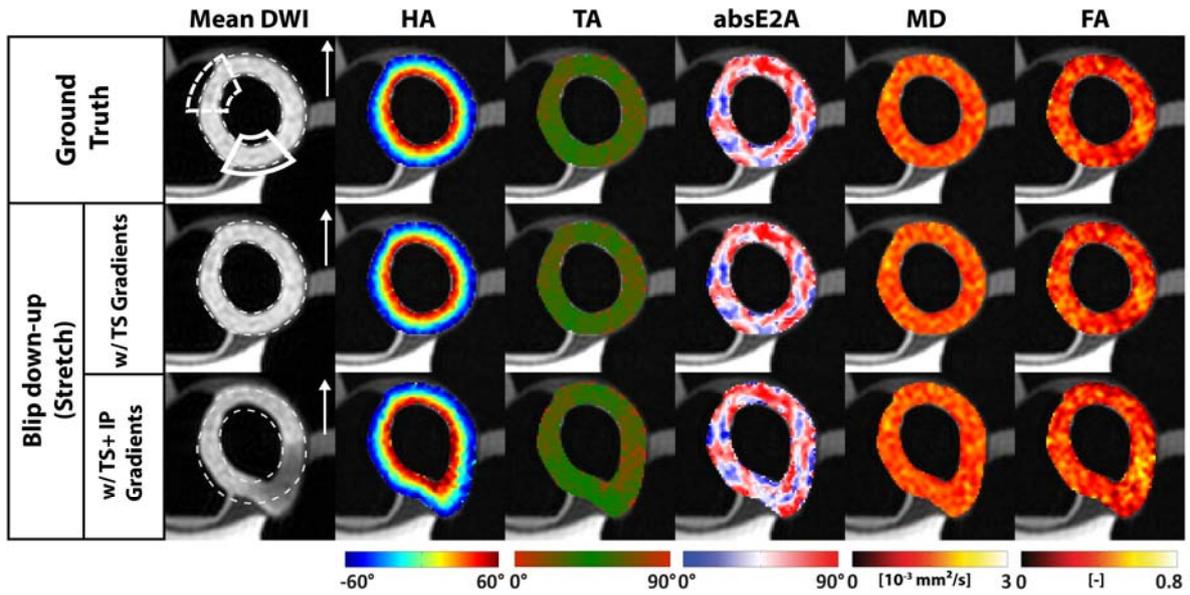

**Figure 6.** MRXCAT-CDTI off-resonance effects on cDTI metrics relative to ground truth (top row), data obtained with through-slice (TS) gradients only (middle row), and data obtained with both through-slice (TS) and in-plane (IP) field gradients (bottom row). Ground truth image contours (white dashed lines) are overlaid. Locations of off-resonance affected area (white solid sector), and remote area (white dashed sector) are shown in the ground truth images. White arrows indicate the phase-encode axis, HA: helix angle, TA: transverse angle, absE2A: absolute sheetlet angle, MD: mean diffusivity, FA: fractional anisotropy.

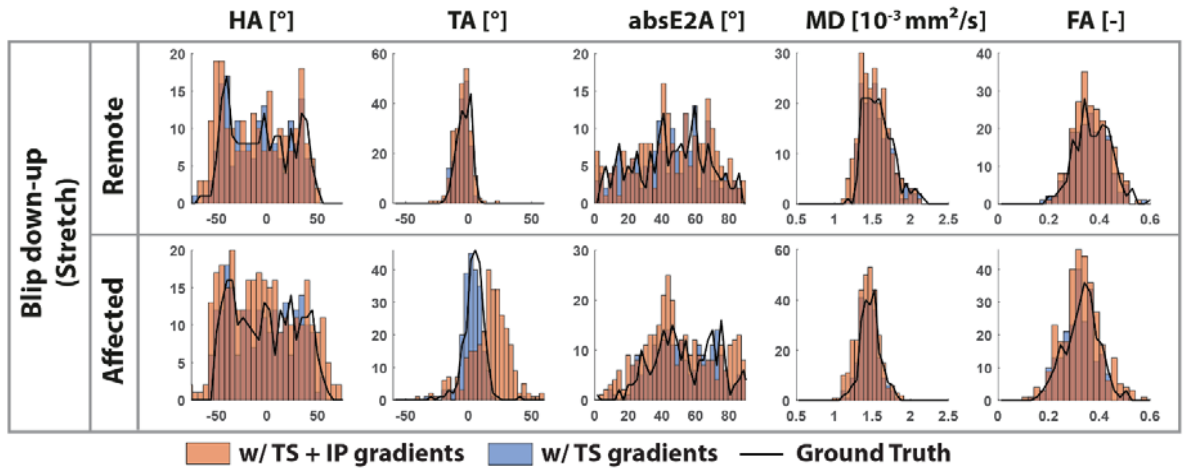

**Figure 7.** Histograms of MRXCAT-CDTI off-resonance example data (from Figure 6) for an off-resonance affected area versus remote area in the presence of only through-slice (TS) gradients (blue), and both through-slice and in-plane (IP) field gradients (red). Each histogram is plotted against ground truth data (solid black line).

Supplementary Materials Table S6 shows histogram intersection values for each MRXCAT-CDTI off-resonance case comparing the remote versus off-resonance affected sector. The average histogram intersection of the remote zone is 94.7% for the through-slice case only, compared to 84.0% for the through-slice and in-plane case; a decrease of 11.3%. The average histogram intersection found in the affected zone is 93.5% for the through-slice only case, whereas in the through-slice and in-plane case it is 74.8%; a 20% decrease. This is most notably caused by the different TA distribution. In the through-slice only case, both sectors have comparable average histogram intersection values between 94% to 95%. For the through-slice and in-plane case the remote sector has an average histogram intersection of 84.0±7.9% compared to 74.8±22.4% for the affected sector; a decrease of 11%.

MRXCAT-CDTI cIVIM lesion example

Fitted parameter maps from a MRXCAT-CDTI cIVIM example with a lesion are shown in Figure 8, with the corresponding distributions for the remote and lesion sectors. The fitted parameter maps for MD, FA, and $f$ display a smoother image appearance, compared to MD*, and FA*. The lesion area is most noticeable in the MD, and $f$ maps. Difference in data distributions between lesion and remote sectors can be best observed for MD, $f$, and MD*. Corresponding mean and SD values of the remote and lesion sectors are summarized in Supplementary Materials Table S7. The lesion-induced influence on the fitted parameters is noticeable for MD, FA, f, MD* and FA* cIVIM metrics.

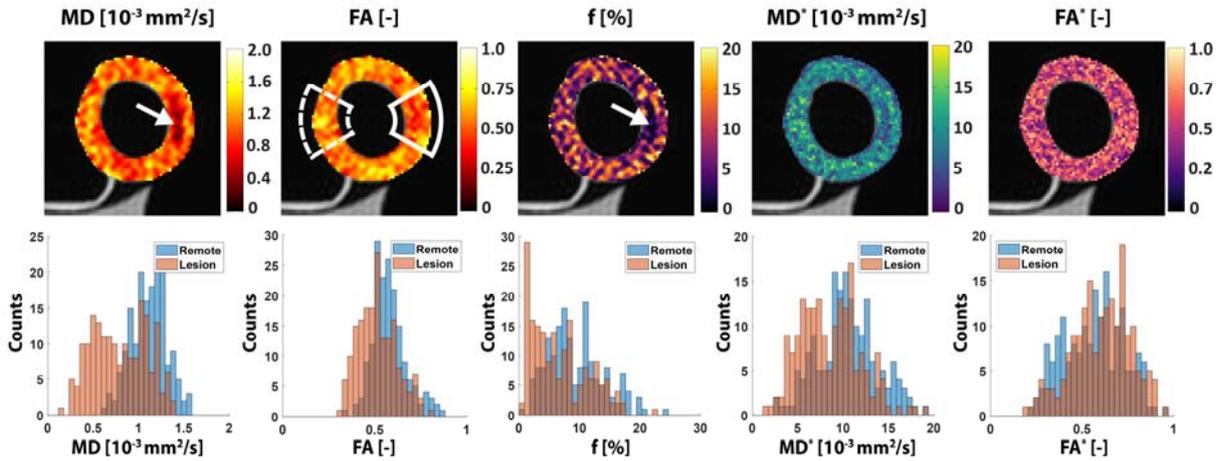

**Figure 8.** MRXCAT-CDTI cIVIM parameters maps. MD and FA maps for both diffusion (D) and perfusion (D*) tensors, as well as the corresponding perfusion fraction map ($f$) (top row). Corresponding histograms for both a remote (blue) and lesion (red) sector (bottom row). Lesion (white solid sector), and remote area (white dashed sector) are shown in the FA map. White arrows indicate lesion section, MD: mean diffusivity, FA: fractional anisotropy, $f$: perfusion fraction.

Computational performance

Total simulation time of MRXCAT-CDTI for the $M_z$ variation examples was 20.8±0.1 minutes for the single-slice cases, and for the free-breathing multi-slice simulation 70.2 minutes. The static off-resonance cases took 9.2±5.4 minutes to simulate, and for the static cIVIM case the simulation time was 12.3 minutes.

# Discussion

In this work a cardiac diffusion and perfusion extension to the numerical MRXCAT framework has been proposed to allow for realistic simulation of image encoding effects providing ground truth data for evaluating image reconstruction methodologies, and quantitative post-processing techniques.

By simulating the finite BW encoding process, partial volume effects, which are present during in vivo MR data encoding, can be simulated. One has to consider that the XCAT software uses non-uniform rational B-splines, which are parametrized using computed tomography data acquired with a spatial resolution of 0.33x0.33x1mm³. Since XCAT can rasterize data to even finer scales, simulating an object at for example 0.25x0.25x0.25mm³ spatial resolution would still be feasible.

The proposed features of MRXCAT-CDTI have been demonstrated for both SE cDTI and STEAM cIVIM diffusion sequences. In the static single-slice SE comparison, applying a correction for the excitation history reduced the average nRMSE across the slice 4.0-fold to 2.6±2.7% for an R-R interval standard deviation of 10%. This highlights the importance of recording and subsequently correcting for R-R interval variations of in vivo measurements when the number of slices and/or heart-rate does not allow full $T_1$ relaxation [11,27].

MRXCAT-CDTI allows for incorporating off-resonances into the image encoding process. Using a synthetic field map, it was shown that MRXCAT-CDTI could replicate the image distortions found under in vivo imaging conditions whilst obtaining comparable off-resonance induced changes in cDTI metrics. Through-slice gradients minimally affected the data as the spatially-induced image shifts and signal dephasing through-slice are considered equal for all images, effectively cancelling its effect out during tensor reconstruction. This leads to high structural similarity and high histogram intersection values compared to ground truth. The presence of in-plane field gradients strongly affected DWI and cDTI metrics. Given the low BW along the phase-encode direction, the in-plane field gradients can either amplify (leading to image stretching) or counteract (leading to image compression) the phase-encode gradients of the EPI trajectory [28,46,47]. Such effects can be retrospectively corrected for in the heart when using a high-resolution field map [28].

The cIVIM example case showed how the introduction of a lesion propagated in the derived metrics compared to an unaffected remote sector.

Differences in HA and absE2A between simulated and in vivo data histograms were found. MRXCAT-CDTI simulates the HA in an idealized manner, and the absE2A map is generated randomly. Therefore MRXCAT-CDTI does not capture all biological variations in in vivo subjects.

With the presented showcases, we have demonstrated the benefit of MRXCAT-CDTI as a tool to study systematic and random influences that occur in a variety of cDTI data encoding scenarios. MRXCAT-CDTI is not limited to these showcases as users can adjust any of the parameters or extend the model by own methods. The existing cine and perfusion MRXCAT implementations can easily be modified to make use of the newer encoding, lesion, and noise operators.

The MRXCAT-CDTI framework simulates $k$-space sampling to generate the data, as opposed to the previous MRXCAT framework which assumed instantaneous encoding. The updated encoder still assumes a static object during $k$-space sampling, hence phase modulation and intravoxel dephasing due to bulk motion, tissue strain or incorrect triggering is not simulated. As proposed by Wissmann et al. [41], this could be overcome by simulating individual XCAT masks per single $k$-space line acquisitions. Alternatively, one could incorporate a warping operator which modifies the initial XCAT mask. To this end, displacement fields used in XCAT to generate the non-rigid motion can be extracted, temporally interpolated, and applied accordingly.

Similar to MRXCAT, MRXCAT-CDTI also relies on signal equations that are solutions to the Bloch equations. At the current point, MRXCAT-CDTI does not emulate diffusion behavior or motion tracking in the presence of gradient waveforms to e.g. evaluate dephasing, but could be incorporated based on particle tracking approaches [67]. Currently, MRXCAT-CDTI only considers data encoding as a 2D problem and through-slice effects are incorporated using a multi thin-slice approach. Extension to true 3D encoding is a logical next step.

The encoding procedure can require considerable processing time depending on the higher spatial resolution of the object. To improve performance at the cost of lower simulation accuracy, a lower 3-fold higher spatial resolution could be used along each spatial axis. Moreover, exploiting GPU instead of CPU processing power or the use of pre-compiled C/C++ code (MEX) could offer a considerable reduction in simulation times.

In conclusion, the proposed open-source MRXCAT-CDTI numerical simulation framework allows for realistic cardiac diffusion and perfusion data simulation to study influences of data encoding, reconstruction and post-processing.

## Author Contributions

Conceptualization, RVG, JW, CTS, and SK; Methodology, RVG, JW, CTS, and SK; Software, RVG, JW, WPS, and SK; Validation, RVG, JW, and SK; Formal Analysis, RVG; Investigation, RVG, and SK; Resources, SK; Data Curation, RVG, and JW; Writing – Original Draft Preparation, RVG, and SK; Writing – Review & Editing, RVG, JW, CTS, WPS, and SK; Visualization, RVG; Supervision, SK; Project Administration, RVG, and SK; Funding Acquisition, CTS, and SK.

## Funding

This research was funded by the Swiss National Science Foundation, grant numbers CR23I3_166485 and PZ00P2_174144. The APC was funded by the ETH Zurich library.

## Institutional Review Board Statement

The study was conducted according to the guidelines of the Declaration of Helsinki, and approved by the Ethics Commission of the Canton of Zurich and institutional guidelines of the University and ETH Zurich.

## Informed Consent Statement

Informed consent was obtained from all subjects involved in the study.

## Data Availability Statement

MRXCAT-CDTI and example scripts to simulate the examples shown in this publication are available online via: https://gitlab.ethz.ch/ibt-cmr-public/mrxcat-cdti-public.

## Conflicts of Interest

The authors declare no conflict of interest. The funders had no role in the design of the study; in the collection, analyses, or interpretation of data; in the writing of the manuscript, or in the decision to publish the results.

# Supplementary Materials

## Supplementary Materials Tables

**Table S1.** MRXCAT-CDTI rejection sampling settings for simulated data.

| Simulation case | Tissue type | Tensor type | Trace Support [$\times 10^{-4}$ mm²/s] | FA Support [$10^{-1}$ -] | Eigenvalue 1, 2, 3 Support [$\times 10^{-4}$ mm²/s] |
|---|---|---|---|---|---|
| SE R-R variations | Healthy myocardium | Diffusion | [27.5, 67.5] | [1.5, 6.5] | [0, 30], [0, 30], [0, 30] |
| SE Off-resonance | Healthy myocardium | Diffusion | [27.5, 67.5] | [1.5, 6.5] | [0, 30], [0, 30], [0, 30] |
| STEAM cIVIM | Healthy myocardium | Diffusion | [12.5, 52.5] | [4.5, 9.5] | [0, 30], [0, 30], [0, 30] |
| STEAM cIVIM | Lesion myocardium | Diffusion | [2.5, 30] | [3.5, 6.5] | [0, 30], [0, 30], [0, 30] |
| STEAM cIVIM | Healthy myocardium | Perfusion | [300, 400] | [4, 6] | [0, 100], [100, 300], [100, 300] |
| STEAM cIVIM | Lesion myocardium | Perfusion | [150, 200] | [3, 5] | [0, 100], [50, 250], [50, 250] |

SE: spin-echo, FA: fractional anisotropy, cIVIM: Cardiac intravoxel incoherent motion.

**Table S2.** In vivo and MRXCAT-CDTI cDTI metrics for a multi-slice dataset corrected for the excitation history.

| Slice | Case | HA endocardial [°] | HA epicardial [°] | TA [°] | absE2A [°] | MD [$10^{-4}$ mm²/s] | FA [$10^{-1}$ -] |
|---|---|---|---|---|---|---|---|
| Base | In vivo | 37.4±21.7 | -39.6±28.8 | -5.2±22.9 | 55.9±28.4 | 15.1±2.2 | 3.2±0.9 |
| Base | MRXCAT | 36.6±15.2 | -39.0±17.2 | -2.4±22.4 | 47.5±23.4 | 14.6±1.5 | 3.3±0.9 |
| Mid | In vivo | 28.0±13.9 | -44.3±26.9 | 0.2±23.7 | 53.1±25.7 | 15.4±2.6 | 3.4±1.1 |
| Mid | MRXCAT | 36.3±16.1 | -38.1±16.7 | -0.1±23.0 | 47.9±23.4 | 14.5±1.4 | 3.3±0.9 |
| Apex | In vivo | 34.6±13.9 | -28.6±25.2 | -3.3±23.6 | 51.6±26.0 | 15.3±2.8 | 3.3±1.1 |
| Apex | MRXCAT | 37.5±16.2 | -37.2±25.2 | -0.7±19.4 | 47.5±23.1 | 14.3±1.4 | 3.3±0.8 |

Data are presented as mean ± SD. HA: helix angle, TA: transverse angle, absE2A: absolute sheetlet angle, MD: mean diffusivity, FA: fractional anisotropy.

**Table S3.** MRXCAT-CDTI data distributions for ground truth, excitation history ($M_z$) uncorrected and corrected single-slice data.

| Simulation case | HA Endocardial [°] | HA Epicardial [°] | TA [°] | absE2A [°] | MD [$10^{-4}$ mm²/s] | FA [$10^{-1}$ -] |
|---|---|---|---|---|---|---|
| Ground truth | 42.5±7.0 | -49.6±10.8 | -0.4±15.5 | 53.7±21.0 | 14.6±1.4 | 3.3±0.7 |
| $M_z$ uncorrected | 42.4±6.8 | -49.3±9.6 | -0.2±16.5 | 52.5±21.4 | 14.7±1.4 | 3.2±0.7 |
| $M_z$ corrected | 42.4±7.0 | -49.4±11.0 | -0.5±15.9 | 53.6±21.0 | 14.5±1.4 | 3.2±0.7 |

Data are presented as mean ± SD. HA: helix angle, TA: transverse angle, absE2A: absolute sheetlet angle, MD: mean diffusivity, FA: fractional anisotropy.

**Table S4.** nRMSE values for excitation history ($M_z$) uncorrected and corrected single-slice MRXCAT-CDTI datasets with respect to ground truth data.

| Simulation case | nRMSE HA Endocardial | nRMSE HA Epicardial | nRMSE TA | nRMSE absE2A | nRMSE MD | nRMSE FA | Mean nRMSE |
|---|---|---|---|---|---|---|---|
| $M_z$ uncorrected | 3.6 | 11.4 | 30.9 | 9.5 | 1.3 | 5.4 | 10.3±10.7 |
| $M_z$ corrected | 0.8 | 5.1 | 6.7 | 1.9 | 0.2 | 0.9 | 2.6±2.7 |

Data are presented in [%]. HA: helix angle, TA: transverse angle, absE2A: absolute sheetlet angle, MD: mean diffusivity, FA: fractional anisotropy.

**Table S5.** MRXCAT-CDTI off-resonance data distributions for ground truth, TS, and TS and IP cases for a remote and off-resonance affected sector.

| Simulation case | Sector | HA Endocardial [°] | HA Epicardial [°] | TA [°] | absE2A [°] | MD [$10^{-4}$ mm²/s] | FA [$10^{-1}$ -] |
|---|---|---|---|---|---|---|---|
| Ground truth | Remote | 37.9±5.0 | -49.7±6.6 | -5.6±5.4 | 43.8±21.4 | 15.5±1.9 | 3.6±0.7 |
|  | Affected | 43.8±6.1 | -45.7±5.8 | 2.6±8.0 | 52.7±18.7 | 14.4±1.3 | 3.2±0.7 |
| TS effects only | Remote | 38.1±5.2 | -49.4±6.4 | -5.3±5.4 | 43.7±21.5 | 15.5±1.9 | 3.6±0.7 |
|  | Affected | 43.1±5.9 | -46.2±5.9 | 2.6±8.1 | 52.7±18.6 | 14.5±1.3 | 3.2±0.7 |
| TS + IP effects | Remote | 38.9±5.9 | -54.1±5.9 | -5.0±6.5 | 46.4±22.5 | 15.2±1.9 | 3.7±0.7 |
|  | Affected | 59.1±15.8 | -48.6±9.8 | 18.2±20.4 | 50.2±21.2 | 14.3±1.5 | 3.3±0.8 |

Data are presented as mean ± SD. HA: helix angle, TA: transverse angle, absE2A: absolute sheetlet angle, MD: mean diffusivity, FA: fractional anisotropy.

**Table S6.** Histogram intersection values of MRXCAT-CDTI off-resonance examples, displaying data from remote and off-resonance affected sectors for all simulation cases with respect to ground truth data.

| Simulation case | Sector | HA | TA | absE2A | MD | FA | Mean HI |
|---|---|---|---|---|---|---|---|
| TS effects only | Remote | 96.0 | 97.2 | 88.1 | 96.0 | 96.0 | 94.7±3.7 |
|  | Affected | 91.4 | 96.8 | 88.2 | 96.4 | 94.6 | 93.5±3.6 |
| TS + IP effects | Remote | 77.3 | 90.4 | 73.7 | 88.2 | 90.2 | 84.0±7.9 |
|  | Affected | 83.2 | 35.4 | 78.5 | 88.6 | 88.1 | 74.8±22.4 |

Data are presented in [%] as mean ± SD. HA: helix angle, TA: transverse angle, absE2A: absolute sheetlet angle, MD: mean diffusivity, FA: fractional anisotropy, HI: histogram intersection.

**Table S7.** MRXCAT-CDTI cIVIM example case diffusion and perfusion tensor fit parameters for remote and lesion sectors.

| Sector | MD [$10^{-4}$ mm²/s] | FA [$10^{-1}$ -] | f [%] | MD* [$10^{-4}$ mm²/s] | FA* [$10^{-1}$ -] |
|---|---|---|---|---|---|
| Remote | 11.2±1.9 | 5.8±1.0 | 9.4±4.6 | 106.1±34.5 | 5.6±1.6 |
| Lesion | 7.9±3.0 | 5.2±1.0 | 6.7±4.7 | 85.7±33.2 | 6.1±1.6 |

Data are presented as mean ± SD. MD: mean diffusivity of the diffusion tensor, FA: fractional anisotropy of the diffusion tensor, $f$: perfusion fraction, MD*: mean diffusivity of the perfusion tensor, FA*: fractional anisotropy of the perfusion tensor.

Supplementary Materials Figures

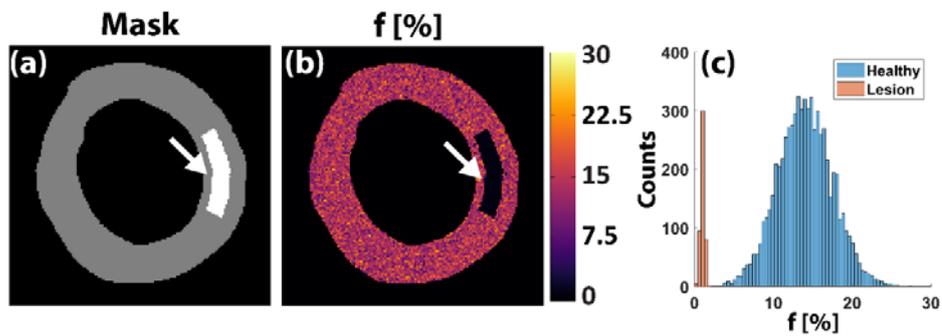

**Figure S1.** MRXCAT-CDTI lesion example. (a) Inside the LV XCAT mask a lesion mask can be defined. (b) This mask can be used to define a map for perfusion fraction $f$. (c) Corresponding distributions for perfusion fraction $f$ of the healthy myocardium and lesion. White arrows indicate the lesion.

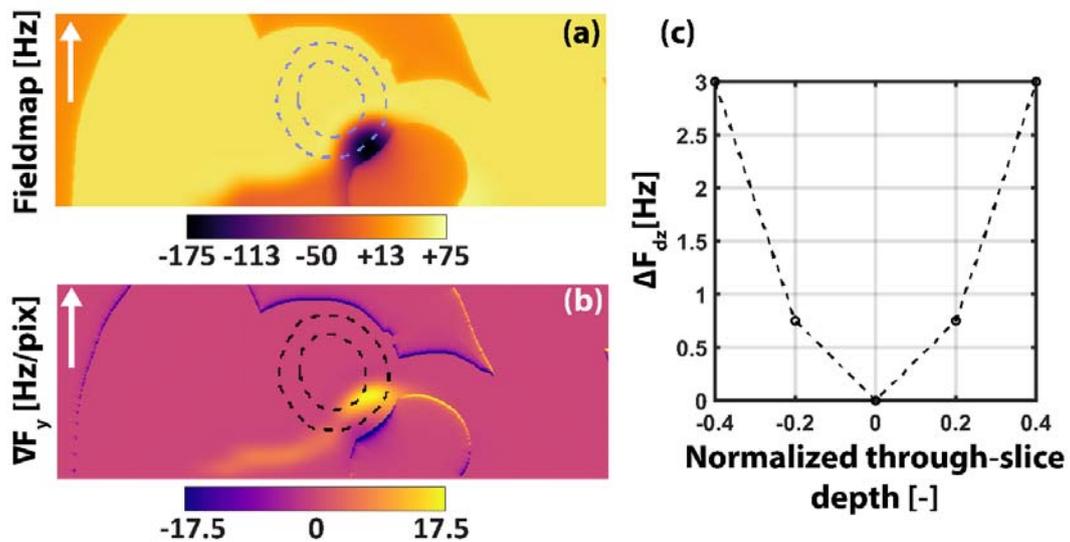

**Figure S2.** Example off-resonance and gradient map, and through-slice offsets for a quadratic frequency profile. (a,b) The field map generated by MRXCAT-CDTI at 0.5x0.5 mm² in-plane resolution with its corresponding gradient map along $F_y$ in [Hz/pixel] at 0.5x0.5 mm² in-plane resolution. (c) A through-slice gradient is achieved by replicating the field map (a) along the through-slice dimension and adding frequency offsets at each $dz$ location. White arrows indicate the phase-encode direction. Contours outline the LV XCAT mask.

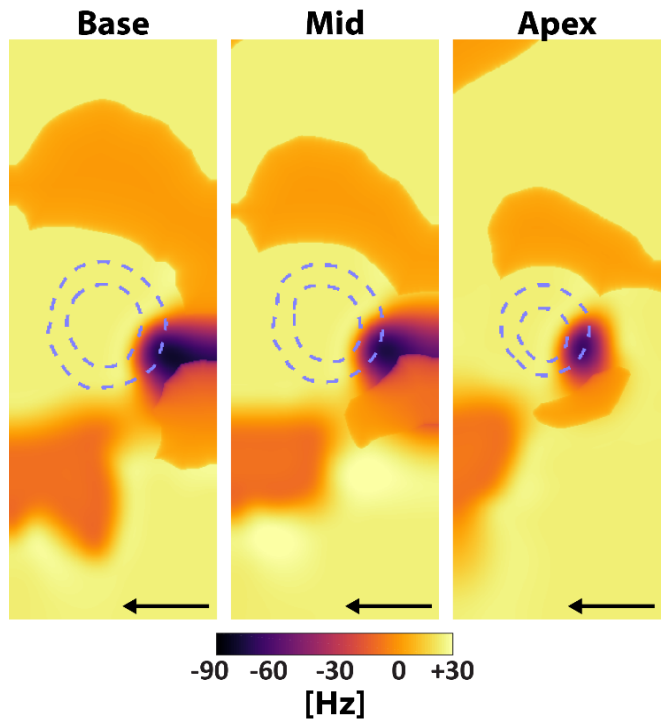

**Figure S3.** Analytical field maps used for the multi-slice MRXCAT-CDTI simulation case. Maps are shown at base-mid-apex short-axis slice positions. Gray dashed lines show the outlines of the left ventricle. Black arrows indicate the phase-encode direction.

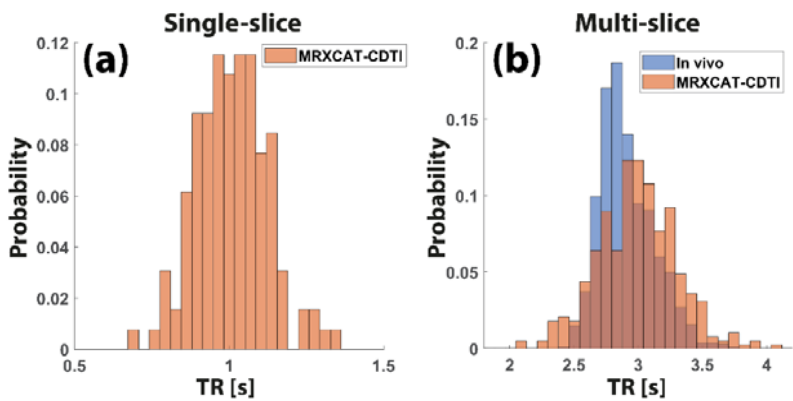

**Figure S4.** Histograms of TR distributions for (a) the single-slice MRXCAT-CDTI excitation history corrected data (red), and (b) in vivo (blue) and MRXCAT-CDTI (red) multi-slice free-breathing data.

# MRXCAT-CDTI Simulation parameters

## General MRXCAT-CDTI simulation parameters

The following general settings were used for the MRXCAT-CDTI simulations: object/$T_2^*$ map/tensor map/field map resolution 0.5x0.5 mm², acquisition resolution in-plane 2.5x2.5 mm² reconstructed to 1.25x1.25 mm², slice thickness = 8 mm, number of coils = 4, $\alpha$ = 90°, FOV = 303x108 mm², and $T_1/T_2/T_2^*$ myocardium = 1000/50/35±5 ms. The presented acquisition BW is for all cases defined at 2.5x2.5 mm² in-plane resolution.

## Multi-slice in vivo vs. MRXCAT-CDTI comparison

A MRXCAT-CDTI dataset at base, mid and apical levels was simulated with the following settings: number of slices = 3, motion states = 10, average myocardial SNR of the $b$ = 0 s/mm² image after including effects of $T_1/T_2/\alpha$ for a single average = 10, signal averages = 10, $T_R/T_E$ = 3000/88 ms, $BW_{epi}$/pixel = 23.4 Hz/pixel / echo-spacing = 1 ms, diffusion gradient scheme with 3 orthogonal directions for $b$ = 100 s/mm² and 9 directions for $b$ = 450 s/mm². Forward model field maps were simulated exhibiting a field gradient of ($\nabla F_y$) of 8 Hz/pixel (at 0.5x0.5 mm² in-plane resolution) at the inferolateral LV heart wall, with a fat frequency offset of -220 Hz, and fat fraction of 10%. An overview of the forward model field maps is provided in Supplementary Materials Figure S3.

Each signal average was randomly assigned to one of 10 available motion state masks to mimic free-breathing respiratory-motion variations. Variations in $M_z$ due to excitation history were introduced by applying an SD on the $T_R$ of 10%.

## MRXCAT-CDTI excitation history correction example

For the excitation history correction example, the following settings were used. A static single short-axis mid-ventricular slice was simulated, motion states = 1, average myocardial SNR of the $b$ = 0 s/mm² image after including effects of $T_1/T_2/\alpha$ = 20, signal averages = 10, $T_R/T_E$ = 1000/88 ms, $BW_{epi}$/pixel = 23.4 Hz/pixel / echo-spacing = 1 ms, EPI blip direction down-up (yielding image stretching) was considered, diffusion gradient scheme with 3 orthogonal directions for $b$ = 100 s/mm² and 9 directions for $b$ = 450 s/mm². Variations in $M_z$ due to excitation history were introduced by applying a standard deviation (SD) on the $T_R$ of 10%.

## MRXCAT-CDTI off-resonance example

For the off-resonance example, the following settings were used. A static single short-axis mid-ventricular slice was simulated, motion states = 1, average myocardial SNR of the $b$ = 0 s/mm² image after including effects of $T_1/T_2/\alpha$ for a single average = 20, signal averages = 10, $T_R/T_E$ = 2000/83 ms, BW$_{epi}$/pixel = 35 Hz/pixel / echo-spacing = 0.7 ms, EPI blip direction down-up (yielding image stretching) was considered, diffusion gradient scheme with 3 orthogonal directions for $b$ = 100 s/mm² and 9 directions for $b$ = 450 s/mm². A ground truth dataset was simulated with no off-resonance effects. The field map was simulated with a field gradient along the y-axis ($\nabla F_y$) of 17.5 Hz/pixel (at 0.5x0.5 mm² in-plane resolution) at the posterior vein location, with a fat frequency offset of -440 Hz, and fat fraction of 10%. The field map was replicated 5-fold along the through-slice direction ($dz$), and a quadratic through-slice frequency profile with mean gradient strength of 7.5 Hz across the slice was added. The field map used in the forward model is shown in Supplementary Materials Figure S2.

## MRXCAT-CDTI cIVIM lesion example

STEAM cIVIM was simulated as follows: single-slice short-axis mid-ventricular static image orientation, average myocardial SNR of the $b$ = 0 s/mm² image after including effects of $T_1/T_2/\alpha$ for a single average = 50, signal averages = 1, $T_R/T_E$ = 2000/35 ms, $\alpha$ = 90°, BW$_{epi}$/pixel = 35 Hz/pixel / echo-spacing = 0.7 ms, diffusion gradient scheme with 6 directions, with $b$-values 0, 25, 50, 75, 100, 150, 200, 250, 300, 400, and 500 s/mm². Perfusion fraction $f$ was set to 0.14±0.04 in the bulk of the myocardium, and to 0.01±0.01 in the lesion area.

Excitation history correction for STEAM data

To perform excitation history correction for STEAM data, the correction factor $K$ at heartbeat index $n$ is calculated according to

$$K(n+2) = \frac{\left(M_0\left(1 - e^{-\frac{T_{R,1}}{T_1(\vec{x})}}\right) + M_z(T_{R,0})\cos(\alpha)e^{-\frac{T_{R,1}}{T_1(\vec{x})}}\right)e^{-\frac{T_{R,2}}{T_1(\vec{x})}}}{\left(M_0\left(1 - e^{-\frac{T_{R,n+1}}{T_1(\vec{x})}}\right) + M_z(T_{R,n})\cos(\alpha)e^{-\frac{T_{R,n+1}}{T_1(\vec{x})}}\right)e^{-\frac{T_{R,n+2}}{T_1(\vec{x})}}}, \qquad \text{(Eq. S1)}$$

with $M_0$ the initial (full) magnetization, $T_{R,n}$ the R-R interval duration of heartbeat $n$, $T_1(\vec{x})$ the local $T_1$ value, $M_z$ the (partially) recovered longitudinal magnetization, and $\alpha$ the flip angle.

When the flip angle $\alpha$ is equal to 90°, Equation S1 can be reduced to

$$K(n+2) = \frac{\left(M_0\left(1 - e^{-\frac{T_{R,1}}{T_1(\vec{x})}}\right)\right)e^{-\frac{T_{R,2}}{T_1(\vec{x})}}}{\left(M_0\left(1 - e^{-\frac{T_{R,n+1}}{T_1(\vec{x})}}\right)\right)e^{-\frac{T_{R,n+2}}{T_1(\vec{x})}}}. \qquad \text{(Eq. S2)}$$